# Chemical bonding in carbide MXene nanosheets

Martin Magnuson[1*], Joseph Halim[1,2] and Lars-Åke Näslund[1]

[1]*Department of Physics, Chemistry and Biology, IFM, Thin Film Physics Division, Linköping University, SE-58183 Linköping, Sweden.*
[2]*Department of Materials Science and Engineering, Drexel University, Philadelphia, PA 19104, USA*

\* Corresponding author: Martin.Magnuson@ifm.liu.se

**Abstract**
The chemical bonding in the carbide core and the surface chemistry in a new group of transition-metal carbides $Ti_{n+1}C_n$-$T_x$ (n=1,2) called MXenes have been investigated by surface-sensitive valence band X-ray photoelectron spectroscopy. Changes in band structures of stacked nano sheets of different thicknesses are analyzed in connection to known hybridization regions of TiC and $TiO_2$ that affect elastic and transport properties. By employing high excitation energy, the photoelectron cross-section for the C *2s* - Ti *3d* hybridization region at the bottom of the valence band is enhanced. As shown in this work, the O *2p* and F *2p* bands are shown to strongly depend both on the bond lengths to the surface groups and the adsorption sites. The effect of surface oxidation and $Ar^+$ sputtering on the electronic structure is also discussed.

## 1. Introduction

In the quest for new 2D materials outperforming graphene [1], research on other more advanced 2D materials has greatly intensified. Despite the large interest in graphene, the lack of a natural band gap and the difficulty of artificially producing one, encourage to explore other 2D materials. Classes of non-graphene 2D materials include hexagonal boron-nitride (BN), molybdenum di-sulphide ($MoS_2$) used as dry lubricant in gears, tungsten di-sulphide ($WS_2$) and *MXenes*, a new family of 2D transition metal carbides denoted $M_{n+1}X_n$, where *M* is a transition metal and X is either carbon or nitrogen [2,3,4]. These complex layered structures that contain more than one element may offer semiconducting and magnetic properties useful for transistors and spintronics. Many other potential applications for MXenes are similar as for doped graphene; 2D-based electronics and screens but on top of that, also energy storage systems such as supercapacitors, Li-ion batteries, fuel and solar cells as well as transparent conductive electrodes and composite materials with high strength.

For MXenes, the parent precursor compounds are inherently nanolaminated materials known as $M_{n+1}AX_n$ (211 for *n*=1, 312 for *n*=2, or 413 for *n*=3) [5] (space group P6$_3$/mmc) phases, where A is a *p*-element that usually belongs to groups IIIA or IVA in the periodic table. These phases contain more than 70 variants, including $Ti_2AlC$ [6] and $Ti_3AlC_2$ [7] that are precursors for the $Ti_2C$ and $Ti_3C_2$ MXenes, respectively. To make MXene from $M_{n+1}AX_n$, the weakly bound and most reactive *A*-layers are etched away and replaced by surface termination groups (-$T_x$) in the exfoliation process [2] where the selection of termination groups, e.g. -O, -OH and –F, influences the material properties. The delamination result in weakly bound stacks of 2D sheets with $M_{n+1}X_n$-$T_x$ composition and larger lattice parameters along the *c*-axis than MAX-phases ($Ti_2AlC$ has *c*=13.6 Å while $Ti_2C$ has *c*=15.04 Å). The parent $M_{n+1}AX_n$ precursors are known to have strong M-X bonds and weak M-A bonds [6-10], but little is known about the electronic structure and chemical bonds of the valence bands in the MXenes. Generally, MXenes consist of a core of a few atoms thick 2D $M_{n+1}X_n$ conductive carbide layers that are crystalline in the basal plane and a transition metal surface that can be functionalized for different properties by changing the chemistry of the termination groups. Complex layered structures like MXenes containing more than one element offer better variation of physical properties than pure materials such as graphene since they can provide a larger number of compositional variables that can be tuned for achieving specific properties. Modeling is important for the understanding of the structure of MXene. Previous works of the electronic structure of MXenes have mainly been based on density functional theory (DFT) calculations [11-13].

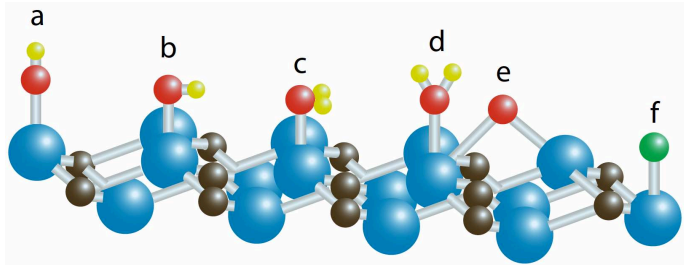

**Figure 1:** Schematic picture of a $M_2C$-$T_x$ structure showing various termination sites of -$T_x$: a) -OH adsorbed on Ti, b) -OH adsorbed on C, c) -O adsorbed on Ti, d) -O adsorbed on C e) a bridged oxygen and e) a -F atom. Blue and black spheres are Ti and C atoms in the $M_2X$ layer, respectively.

In this work, we aim to present experimentally the electronic structure and chemical bonding in the valence bands of $Ti_2C$-$T_x$ and $Ti_3C_2$-$T_x$ MXenes in comparison to TiC as a reference material by using valence band X-ray photoelectron spectroscopy (XPS). By employing X-rays with relatively high energy (Al $K_\alpha$ 1486.6 eV), we enhance the photoelectron cross-section for the hybridization region of the Ti *3d* - C *2s* states at the bottom of the valence band. The XPS valence band spectra are modeled by different structure components by electronic structure calculations to explore the sensitivity to various adsorption sites for different -O, -OH and -F surface groups as schematically shown in Fig. 1. For detailed electronic structure studies of termination groups, the necessity of clean non-oxidized surfaces is discussed.

## 2. Experimental details

### 2.1 Sample preparation
Flakes of $Ti_2C$-$T_x$ were produced by immersion of $Ti_2AlC$ powders (3-ONE-2, Voorhees, NJ, > 92 wt. % purity) of particle size of < 35 μm in 10 % conc. Hydrofluoric acid (HF(aq)) (Fisher Scientific, Fair Lawn, NJ) for 10 h at room temperature. For more details, see Refs. [2-4,14]. Similarly, flakes of $Ti_3C_2T_x$ were produced by immersion of $Ti_3AlC_2$ powders in 50% conc. HF(aq) for 18 h at room temperature. The $Ti_3AlC_2$ powders were prepared by mixing commercial $Ti_2AlC$ powders (Kanthal, Sweden, -325 mesh) with TiC (Alfa Aesar, Ward Hill, USA, 99.5 wt.% purity, -325 mesh) in a 1:1 molar ratio (after adjusting for the ~ 12 wt.% $Ti_3AlC_2$ already present in the commercial powder), followed by ball milling for 18 h to ensure mixing of the powders. The mixture was then heated in a tube furnace to 1350 °C with a rate of 10 °C min$^{-1}$ for 2 h under flowing argon (Ar).

After the HF treatment, the etched powders were washed in cycles of deionized water, centrifugation, decanting, followed by filtration using a vacuum assisted filtering device to dry the powder from the water or suspension in ethanol and dried by evaporation to yield the MXene powders. The dried powders were then cold pressed using a load corresponding to a stress of 1 GPa in a steel die to produce freestanding discs that were used in this study.

The TiC sample was made from powder (Alfa Aesar, Ward Hill, USA, 99.5 wt.% purity, -325 mesh) that was cold pressed, performed similar to the $Ti_3C_2$ and $Ti_2C$.

**2.2 Valence band XPS measurements**
The valence band XPS measurements were performed using a surface analysis system (Kratos AXIS Ultra$^{DLD}$, Manchester, U.K.) using monochromatic Al-K$_\alpha$ (1486.6 eV) radiation. The X-ray beam irradiated the surface of the samples at an angle of 45°, with respect to the surface, and provided an X-ray spot size of 300 x 800 μm. The electron energy analyzer accepted the photoelectrons perpendicular to the sample surface with an acceptance angle of ± 15°. XPS spectra were recorded for the valence bands with a step size of 0.1 eV and a pass energy of 20 eV, which provided an overall energy resolution better than 0.5 eV. The binding energy scale of all XPS spectra was referenced to the Fermi-edge ($E_f$), which was set to a binding energy of zero eV. The MXenes in this work were examined before and after low angle incident (20°) Ar$^+$ sputtering for 600 s using a 4 keV Ar$^+$ beam raster of 2x2 mm$^2$ over the probed area.

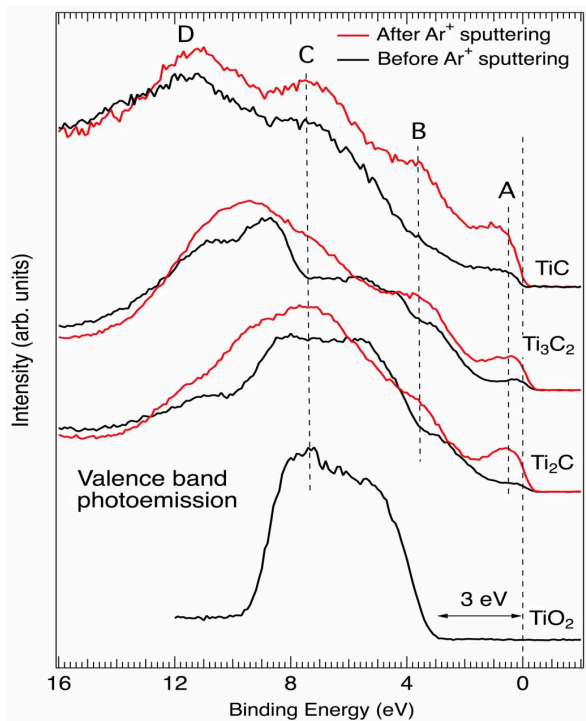

**Figure 2:** Valence-band XPS spectra of $Ti_2C-T_x$ and $Ti_3C_2-T_x$, before and after Ar+ sputtering, in comparison to TiC and amorphous $TiO_2$. The vertical dashed line at 0 eV represents the Fermi level ($E_F$) on the binding energy scale ($E_b$).

**2.3 DFT calculations**
First principle calculations were carried out using density functional theory (DFT) implemented in the Vienna *ab initio* simulation package (VASP) [15]. The projected augmented wave (PAW) [16] method and the Perdue-Burke-Ernzerhof (PBE) functional of the generalized gradient approximation (GGA) [17] with a *k*-point grid of 21x21x3 was used in the calculations. Figure 1 illustrates an $M_2C-T_x$ structure showing various -$T_x$ termination sites positioned on top of Ti, on to of C, and in a Ti-Ti bridged position.

## 3. Results

Figure 2 shows valence-band XPS spectra of $Ti_2C–T_x$ and $Ti_3C_2–T_x$ before and after sputtering in comparison to TiC and amorphous anatase-like $TiO_2$. The benefit of using 1487 eV photon energy is obvious, i.e. the structures at the bottom of the band are dominating the spectra; the drawback is, however, the broadening of the structures. The valence band structure of TiC, obtained using 1487 eV photon energy, has been presented previously [20] and a direct comparison with that work shows that the structures A at 0.5 eV, B at 3.5 eV, and D at 10.9 eV are associated with the metallic Ti *3d* − Ti *3d* hybridization, the Ti *3d* - C *2p* hybridization, and the C *2s* states, respectively [18,20,21]. These three structures are accompanied by a fourth structure at 7.5 eV, denoted by C in Figure 2, which does not originate from TiC. A comparison with the valence band structure of pure $TiO_2$ suggests that C is associated with the O *2p* states of oxidized Ti. The intensity of all structures is enhanced after $Ar^+$ sputtering. Even the O *2p* structure is enhanced, which indicates that $TiO_2$ is still present although the O *1s* XPS indicates that the surface oxide is removed. However, as the valence band XPS has a larger probing depth than the O *1s* XPS, the oxygen contribution in the VB spectra (after sputtering) originates from atomic layers below the surface that are not detectable by O *1s* XPS. Moreover, for the photoemission energy of the valence band spectra, oxygen has three times higher photoionization cross-section than carbon. Thus, for TiC, the origin of peak C after sputtering is associated with the relatively large sensitivity of oxygen in combination with the large probe depth.

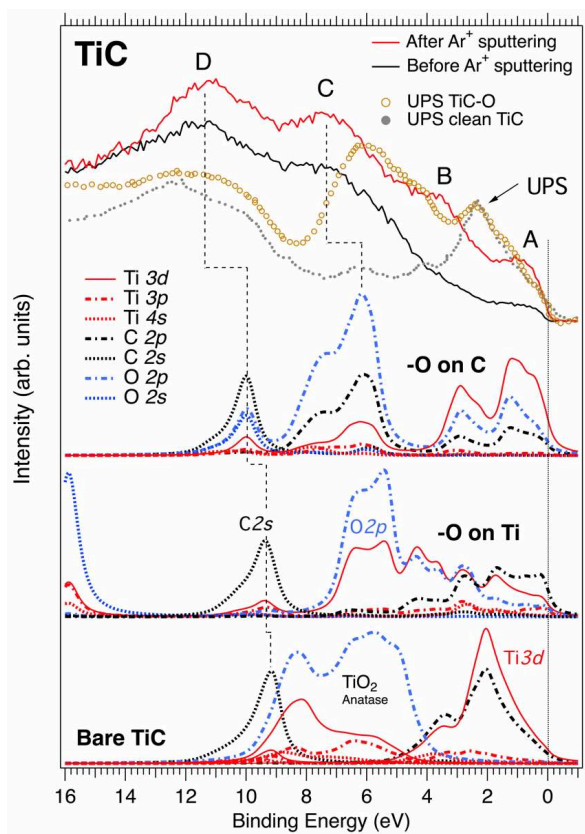

**Figure 3:** Valence band XPS spectra of TiC before and after $Ar^+$ sputtering of the surface compared to UPS data [18] before and after oxygenation. DFT calculations with -O adsorption on top of the Ti and C sites are shown below together with bare TiC and $TiO_2$ oxide in the form of anatase.

The dashed lines highlight the trend in the peak structures A-D observed for the TiC and the corresponding valence band structures shown in the spectra for $Ti_2C-T_x$ and $Ti_3C_2-T_x$. As observed, bands A, B, and C remain at the same binding energies for all carbon-containing samples. The intensity of the peak at ~0.5 eV, corresponding to the Ti *3d* – Ti *3d* orbital overlaps, is enhanced after $Ar^+$ sputtering. The intensity at the $E_F$ is significantly smaller for TiC and is consistent with the higher resistivity in comparison to both $Ti_2C$ and the more carbidic $Ti_3C_2$ [6,7]. Also band B at 3.7 eV appears more clearly after surface cleaning. The position of band C coincides with the main peak at 7.5 eV of amorphous $TiO_2$. The bandgap of the insulator $TiO_2$ in the form of amorphous anatase is ~3.0 eV as indicated by the horizontal arrow at the bottom.

For $Ti_2C$ and $Ti_3C_2$, the spectra before sputtering (black curves) exhibit a plateau with high intensity in the region corresponding to amorphous $TiO_2$ that is mainly caused by oxidation of the surfaces. After sputtering (red curves), the intensity corresponding to the carbide structures becomes clearer, however, the signal from the surface oxide is not reduced because of the reasons mentioned above. Contrary to bands A-C, the C *2s* band D at the bottom of the valence band exhibit dispersion from the $E_f$ as the C-content increases. The energy shift of the C *2s* states towards the $E_f$ indicates different Ti-C bonding in the MXenes than in the reference sample TiC. Furthermore, the Ti-C bond strength in $Ti_2C-T_x$ appears to be different than in $Ti_3C_2-T_x$, which probably is because of thinner layers and additional interfaces. The valence band spectra of $Ti_2C-T_x$ and $Ti_3C_2-T_x$ show, in addition to bands A-D, structures around 4.5, 5.6, and 8.9 eV, which probably originate from the termination groups on the $Ti_2C$ and $Ti_3C_2$ surfaces. In order to obtain a deeper understanding and interpretation of the nature of the spectral band features and the influence of termination groups, we compare the measured valence band spectra to calculations with different components.

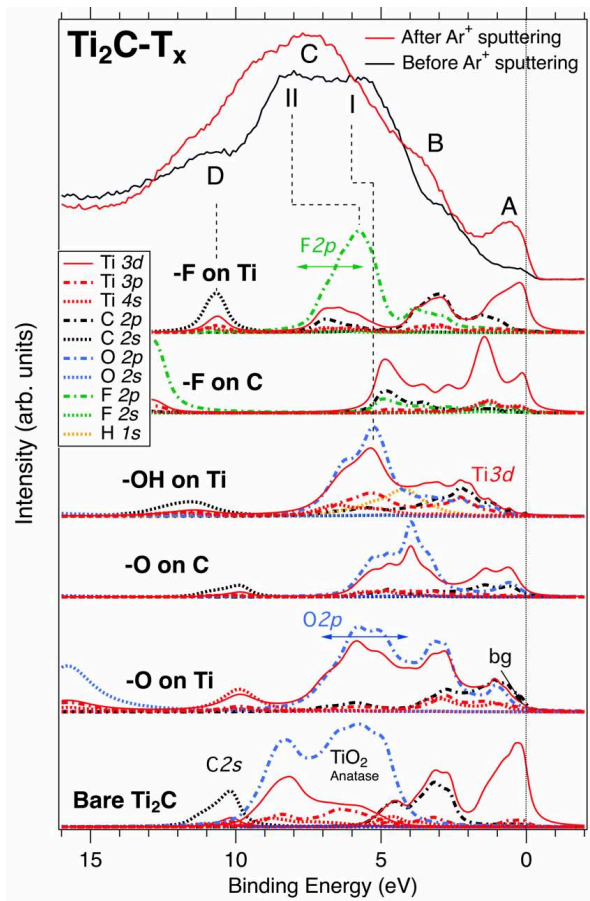

**Figure 4:** Valence band XPS spectra of $Ti_2C-T_x$ before and after $Ar^+$ sputtering of the surface. DFT calculations with -O, -OH and -F termination groups at different adsorption sites (O on Ti and C as well as bridge position) and bare $Ti_2C$ and $TiO_2$ oxide in the form of anatase are shown below.

Figure 3 shows the reference carbide material TiC together with modeled spectra using DFT calculations as described in section 2.3. Ultraviolet photoemission spectra (UPS) of bare and -O covered TiC are also shown for reference at the top (dotted curves) [18,20,21]. Generally, the calculated peaks are in good agreement with the experimental spectra. For comparison, the UPS spectra before and after oxygenation [18] of clean TiC are well reproduced by the calculations at the bottom of Fig. 3. UPS has a photoemission cross-section more suitable for probing the Ti *3d* states close to the $E_f$. Generally, a high density of states (DOS) at the Fermi level ($E_F$) is correlated to low resistivity and we note that there are states at the $E_f$ indicating that TiC is a conductor (resistivity 2.5 μΩm [6]), although there are also other issues that affect the transport properties [3,19]. Adsorption of oxygen at the TiC surface gives rise to a peak at ~6 eV in the UPS spectrum. However, the adsorption site of oxygen on a TiC surface is known to depend on the surface cutting (100 or 111) [19,20]. For comparison, N on a TiN surface prefers to adsorb on top of N rather than on top of Ti [21]. Our DFT calculations show that O adsorbed on top of Ti would give rise to a band at 16 eV due to Ti *3d* – O *2s* hybridization and this band would not occur in the case of O adsorbed on C. As the XPS spectra in Figure 3 do not exhibit this band we conclude that oxygen probably does not adsorb on top of Ti for TiC. It is also observed that the O *2p* band position in the calculations changes between 6-8 eV depending on the bond length between the surface and the adsorbate.

Figure 4 shows $Ti_2C-T_x$ MXene together with calculated spectra with different termination groups and adsorption sites. Bare $Ti_2C$ is predicted to be metallic with more Ti *3d* states at $E_f$ than $Ti_2AlC$ [12]. Peaks that are associated with bands from –O and –F termination species are indicated with I and II, respectively. When –F and –OH surface termination groups are included in the calculations, new bands are formed. Generally, the calculations predict that there are states at $E_f$ except for pure –O termination that gives a semiconductor with a bandgap of ~0.24 eV [12], as indicated in Fig. 4. Experimentally, $Ti_2C-T_x$ has a resistivity of 68 μΩ·m [3,22,23].

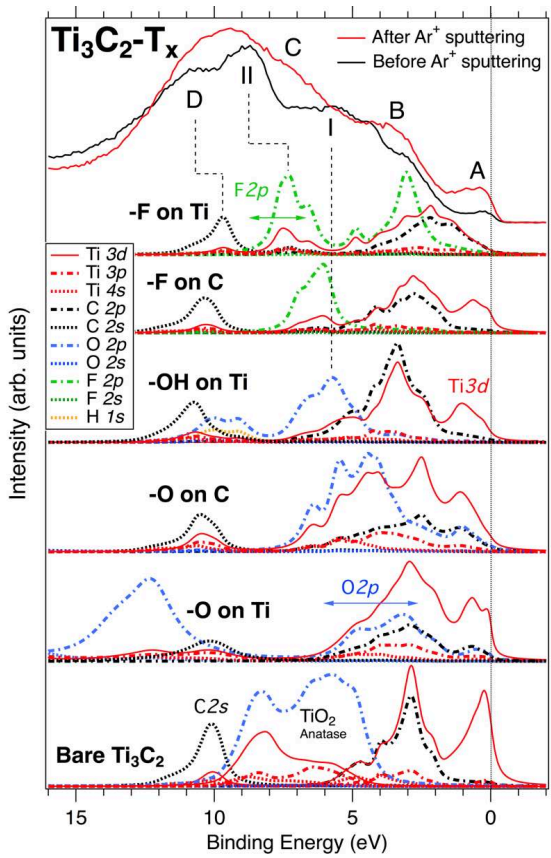

**Figure 5:** Valence band XPS spectra of $Ti_3C_2-T_x$ before and after $Ar^+$ sputtering of the surface. DFT calculations with -O, -OH and -F termination groups at different adsorption sites (O on Ti and C as well as bridge position) and bare $Ti_3C_2$ and $TiO_2$ oxide in the form of anatase are shown below.

Bands A and B are essentially the same as for TiC. Theoretically, the positions of bands I (5-7 eV) and II (8-10 eV) corresponding to O *2p* and F *2p* states, respectively, strongly depend on both the adsorption site and the bond length to the $–T_x$ species.

For the calculated spectra, the horizontal arrows indicate the relatively large span of band positions that occur for different $Ti-T_x$ bond lengths. Generally, a longer (weaker) $Ti-T_x$ bond length for a particular structure results in a shorter energy distance from the $E_F$ [7,8,19]. The calculated (relaxed) Ti-O bond length (~1.99 Å) is shorter than the Ti-F bond length (~2.15 Å). This is probably related to the difference in the number of $2p^n$ valence electrons and the difference in electronegativity between O and F that causes smaller charge-transfer to O (~1e$^-$) in comparison to F (~2e$^-$) from $Ti_2C$ [11]. Theoretically, the absolute energy positions of the O and F bands also depend on the choice of exchange correlation functional [12] in the DFT calculations.

Figure 5 shows $Ti_3C_2-T_x$ MXene together with calculated spectra with different adsorption sites and termination groups. Peaks that are associated with bands from –O

and –F terminations are indicated with I and II, respectively. As in the case of $Ti_2C$-$T_x$, the DFT calculations predict bare $Ti_3C_2$ MXene to be metallic with finite Ti *3d* states at $E_f$. The DOS at $E_f$ for bare $Ti_3C_2$ is higher than for the $Ti_3AlC_2$ parent MAX-phase. Theoretically, the DOS at $E_f$ increases with *n* [12] and for $Ti_3C_2$-$T_x$, there are always states at $E_f$, even for pure –O surface groups. Generally, the metallic properties weaken with increasing *n* as more Ti-C bonds are formed. Experimentally, the resistivity of the $Ti_3C_2$-$T_x$ sample is ~5 μΩm [3,23].

Band I is the O *2p* states and is located at 5-7 eV. The Ti-O hybridization appears to be weaker for $Ti_3C_2$-O than in the case of $Ti_2C$-O based on intensity differences. Band II is associated with F *2p* states and is located at 8-10 eV. As in the case of $Ti_2C$-$T_x$ MXene, the calculated positions of the O *2p* and F *2p* bands strongly depend both on the –$T_x$ bond length and the absorption site. Recently, calculated predictions have indicated that the surface groups are more likely located in hollow adsorption sites [11-13]. However, the exact locations of the surface groups are complicated and largely depend of the structure *n* and elements involved as well as temperature. Based on the difference in energy positions from the $E_F$, the -F termination groups appear to be differently bonded to the surface than the –O termination species.

## 4. Discussion

Several interesting observations are obtained from the valence band XPS spectra shown in Figs. 2-5. The weight of the peaks and their energy positions give information about the hybridization regions in the strength of the chemical bonds between the containing elements that affect mechanical properties such as ductility and elasticity. The valence band XPS data show intensity redistributions of the C *2s* states between 9 and 11 eV at the bottom of the valence band. The intensity redistributions might suggest a weaker Ti-C bond strength in $Ti_2C$-$T_x$ than in $Ti_3C_2$-$T_x$ and TiC. However, in the calculations, the C *2s* peak appear at 9.2-10 eV for TiC, at 10.1-10.7 eV for $Ti_3C_2$ and 10.2-10.9 eV for $Ti_2C$.

Generally, a weaker bond in one direction is compensated by a stronger bond in another direction [8,19]. This has previously been observed by strengthening of the Ti-C bond as the Al-layer is inserted in the MAX-phases e.g., $Ti_2AlC$ [6], $Ti_3AlC_2$ [7]. A strengthening of the covalent Ti-C bond in MAX-phases compared to TiC was also theoretically confirmed by balanced overlap population analysis [9,10], where the distance from the $E_f$ and the intensity of the overlap determines the strength of the bond. For example, the Ti-Al overlap appears closer to the $E_f$ in comparison to Ti-Si and Ti-Ge overlaps, signifying a weaker covalent bond. In analogy, the bare MXenes $Ti_2C$ and $Ti_3C_2$ without surface groups would have stronger Ti-C bonds than the corresponding MAX-phases $Ti_2AlC$ (2.102 Å), $Ti_3AlC_2$ (2.086 Å) and TiC (2.164 Å). However, surface groups of MXenes withdraws charge from the Ti-C bonds of the conductive carbide core and, thus, weakens these bonds, which affect the C *2s* - Ti *3d* hybridization region at the bottom of the valence band. As shown by the intensity redistributions in the C *2s* - Ti *3d* hybridization region towards the $E_f$, the Ti-C bond is lengthened in $Ti_2C$-$T_x$ compared to $Ti_3C_2$-$T_x$. This may affect both the transport and elastic properties of the MXene materials.

More interestingly, our DFT calculations show that the positions of the O *2p* and F *2p* bands strongly depend both on the adsorption sites and bond lengths to the surface groups. This implies that the bond strength can be modified to optimize the elasticity. Moreover, moving the bands is useful to tailor-make the catalytic ability and surface oxidation resistance for different MXenes by the choice of transition elements and termination species. Ti in $Ti_2C$ likely has a higher oxidation state than Ti in $Ti_3C_2$ due to more interfaces and functional groups (-OH, -O and -F). Theoretical modeling of different adsorption sites show that adsorption sites close to Ti gives higher oxidation states for Ti than for adsorption sites close to C. However, the surface groups may also adsorb in bridge positions or with bonds in an angle relative to the surface plane rather than straight on top of Ti and C [13].

Generally, the calculated spectra on top of Ti than on top of C in Fig. 4 and 5 appear to be in better agreement with the experimental data. Thus, the peak structures in the calculations indicate that adsorption on top of Ti may be more likely than adsorption on top of C. However, there may be other more energy favorable adsorption sites such as bridge and hollow sites that. Recent ground state calculations (neglecting temperature effects such as phonons) indicate that -F may first adsorbs in a hollow position in a FCC arrangement vis-à-vis the other atoms in the structure [11-13], while -O is more "itinerant" and may adsorb either on the same hollow site position if it is empty from –F or otherwise at other adsorption sites. Based on these predictions, other adsorption sites than on-top sites, such as bridge and hollow sites should be worth further investigation.

In MAX-phases, like in TiC and TiN, there are metal bonds and covalent bonds with ionic character that depend on the electronegativity of the constituent elements. As in the case of MAX-phases, the chemical bonding in the conductive carbide core of the MXenes is dominated by covalent bonds. However, the transition metal oxide surface has more ionic character in the chemical bonds. By changing the surface groups, it may become possible to modify the bond strength by changing the ionicity and tailor these materials with desired macroscopic properties such as conductivity and magnetism [19]. Generally, a high density of states (DOS) at the Fermi level ($E_F$) is correlated with low resistivity but there are also other issues that affect the transport properties [19]. In particular, the resistivity is higher in $Ti_2C$ (68$\mu\Omega \cdot m$) than in $Ti_3C_2$ (5$\mu\Omega \cdot m$) [23] but the DOS is similar at the $E_F$ and calls for deeper structural analysis. Contrary to bare MXene that is metallic, terminated MXene may become a narrow-band gap semiconductor material. It is therefore of fundamental interest for the development of semiconductor components to be able to tune the electronic properties to achieve a direct band gap.

Generally, MXenes are explored for many potential applications such as anodes in lithium-ion batteries, supercapacitors and in catalysis [2,3,4]. This is because of the 2D nature of the material with a large surface area and good thermal conductivity [24] in the conductive carbide core of the MXenes. Furthermore, according to the "*d*-band center theory" [25], the valence band maximum can be modified to optimize the catalytic ability for different MXenes. In this way, the electronic band filling and thus the position of the *3d* or *4d*-band can be moved by the choice of transition metal. Moreover, the surface chemistry of MXenes can be changed by using different termination species to optimize different surface reactions. Here, we point out the possibilities of influencing the *d*-band structure by the choice of layer thickness, *n*. In addition, the choice of transition elements in the conductive carbide core of the MXene in combination with a choice of different termination species. Up to now, only the O/F ratio can be tuned in the synthesis process by changing the etching parameters such as acid, concentration and etching time on MAX-phases. However, it is not known how to freely tune termination species or how to produce bare or pure MXene with only one type of termination species.

However, all aged carbide samples may contain $TiO_2$ on the surface that largely affect the electronic structure and make valence band XPS studies difficult. In addition, $Ar^+$ sputtering of the samples will remove termination groups and simply damage or destroy the surface of the studies MXene samples. Furthermore, a decreasing oxygen signal at the surface was monitored by the O *1s* XPS peak. Still the valence band spectra suggest that there is a significant contribution from oxidized Ti. The reason suggests being that as the valence band XPS spectra have a larger probe depth (~20 Å) than the O *1s* XPS spectra (~15 Å) and the plane distances in MXenes are about 10 Å, the oxygen contribution in the valence band spectra (after sputtering) originates from atomic layers below the surface that are not detectable by O *1s* XPS. Moreover, for the photoemission energy of the valence band spectra, oxygen has three times higher photoionization cross-section than carbon. Thus, for TiC, the origin of "band C" after sputtering is because of the higher sensitivity of oxygen in combination with the large probe depth.

Although this study was made on aged cold-pressed powder samples to emphasize problems associated with oxidation of MXene, interesting information is obtained by valence band XPS. The higher intensity for $Ti_3C_2$ MXene at ~9 eV can be associated with a higher fraction of the -F termination in comparison to $Ti_2C$ MXene. This is consistent with core-level XPS studies that provide information about relative fractions of the termination groups [14]. However, more systematic spectroscopy measurements in comparison to calculated spectra is needed to clarify this issue. Future studies may involve *in-situ* preparation of thin-film samples to avoid surface oxides to more clearly isolate spectrum features related to the termination groups. From a computational point of view, we also emphasize a suitable choice of exchange correlation functional to model band gaps and realistic peak structures close to the $E_F$. Structural models often include one termination species while in reality, there is a mixture of termination species that may have a combined effect on stability and properties. Future more advanced models may include photoemission matrix elements and cross section effects as well as temperature effects and phonon stabilization.

Thus, there are many remaining challenges for MXene to be useful in potential applications. We anticipate that spectroscopies such as X-ray absorption, resonant photoemission and angular-resolved photoemission spectroscopy (ARPES) will be useful tools to identify the details of the bonding of the surface groups at the interfaces. For TiC surfaces, the photoionization cross section vary considerably with incident photon energy as best observed using synchrotron radiation. By taking advantage of polarized and tunable synchrotron radiation, a better insight into the origin of the atomic orbitals through detailed energy-dependent cross section and depth analysis may be obtained.

## 5. Conclusions

Valence-band X-ray photoelectron spectroscopy of $Ti_{n+1}X_n$-$T_x$ MXenes can be used to probe changes in Ti-C bond strength of the conductive carbide core that depend on the carbon content and atomic layer thickness. The effect of the termination groups and especially how they withdraw charge from the Ti-C carbide bonds in the conductive core layers can also be studied. As shown, the energy positions of the O *2p* (~6 eV) and F *2p* (~9 eV) bands strongly depend both on the adsorption sites and the bond lengths to the surface groups. However, a prerequisite for a qualitative study is that the samples are free from oxidized material, *i.e.* the samples must be freshly made in operando, and $Ar^+$ sputtering must be avoided.

## 6. Acknowledgements
We would like to thank M. Barsoum for fruitful discussions. This work was supported by the Swedish Research Council (VR) Linnaeus Grant LiLi-NFM, the FUNCASE project supported by the Swedish Strategic Research Foundation (SSF). MM acknowledges financial support from the Swedish Energy Research (no. 43606-1), *the Swedish Foundation for Strategic Research (SSF)* (no. RMA11-0029) *through the synergy grant FUNCASE* and the Carl Tryggers Foundation (CTS16:303, CTS14:310). We would like to express our thanks to the reviewers of the present work for very insightful and helpful comments.